\pdfoutput=1  
\documentclass[]{article}  
\usepackage{url,float}
\usepackage{graphicx}
\usepackage{amsmath}
\usepackage{amsfonts}
\usepackage{amssymb}
\usepackage{latexsym}

\newcommand{\hide}[1]{}

\newcommand{\ABox}{
\raisebox{3pt}{\framebox[6pt]{\rule{6pt}{0pt}}}
}
\newenvironment{proof}{{\bf Proof:}}{\hfill\ABox}

\newtheorem{theorem}{{\bf Theorem}}

\newtheorem{lemma}{Lemma}

\newcommand{\lemlab}[1]{\label{lemma:#1}}
\newcommand{\thmlab}[1]{\label{thm:#1}}

\newcommand{\figlab}[1]{\label{fig:#1}}
\newcommand{\seclab}[1]{\label{sec:#1}}
\newcommand{\secref}[1]{\ref{sec:#1}}
\newcommand{\figref}[1]{\ref{fig:#1}}

{\makeatletter
 \gdef\xxxmark{%
   \expandafter\ifx\csname @mpargs\endcsname\relax 
     \expandafter\ifx\csname @captype\endcsname\relax 
       \marginpar{xxx}
     \else
       xxx 
     \fi
   \else
     xxx 
   \fi}
 \gdef\xxx{\@ifnextchar[\xxx@lab\xxx@nolab}
 \long\gdef\xxx@lab[#1]#2{{\bf [\xxxmark #2 ---{\sc #1}]}}
 \long\gdef\xxx@nolab#1{{\bf [\xxxmark #1]}}
 \gdef\turnoffxxx{\long\gdef\xxx@lab[##1]##2{}\long\gdef\xxx@nolab##1{}}%
}

\def\a{{\alpha}}
\def\b{{\beta}}

\def\R{{\mathbb{R}}}

\title{%
Flat Zipper-Unfolding Pairs
for Platonic Solids
} 

\author{%
Joseph O'Rourke%
    \thanks{Department of Computer Science, Smith College, Northampton, MA
      01063, USA.
      \protect\url{orourke@cs.smith.edu}.}
}

\begin{document}
\maketitle

\begin{abstract}
We show that four of the five Platonic solids' surfaces may be cut open
with
a Hamiltonian path along edges and unfolded to a polygonal net
each of which can ``zipper-refold'' to a flat doubly covered
parallelogram,
forming a rather
compact representation of the surface.
Thus these regular polyhedra have particular flat ``zipper pairs.''
No such zipper pair exists for a dodecahedron, whose Hamiltonian
unfoldings are ``zip-rigid.''
This report is primarily an inventory of the possibilities, and raises
more questions than it answers.
\end{abstract}

\section{Introduction}
\seclab{Introduction}
It has been known since the time of Alexandrov---and it was certainly
known to him---that the surface of a polyhedron could sometimes be cut
open to a net and refolded to a doubly covered polygon, which we will
henceforth call a \emph{flat polyhedron}.
Such flat polyhedra are explicitly countenanced in Alexandrov's 1941
gluing theorem.\footnote{
See~\cite[Sec.~23.3]{do-gfalop-07}
and~\cite[Sec.~37]{p-ldpg-10} for descriptions of this theorem.
}
Perhaps the first specific example of this possibility occurred in~\cite{lo-wcpfp-96},
which included the example illustrated in
Figure~\figref{LatinCrossParallelogram}:
the familiar Latin-cross unfolding of the cube may be refolded to a
flat convex quadrilateral polyhedron.
This is one of the two flat convex polyhedron that may be folded
from
the Latin cross~\cite[Fig.~25.32]{do-gfalop-07}.
\begin{figure}[htbp]
\centering
\includegraphics[width=0.8\linewidth]{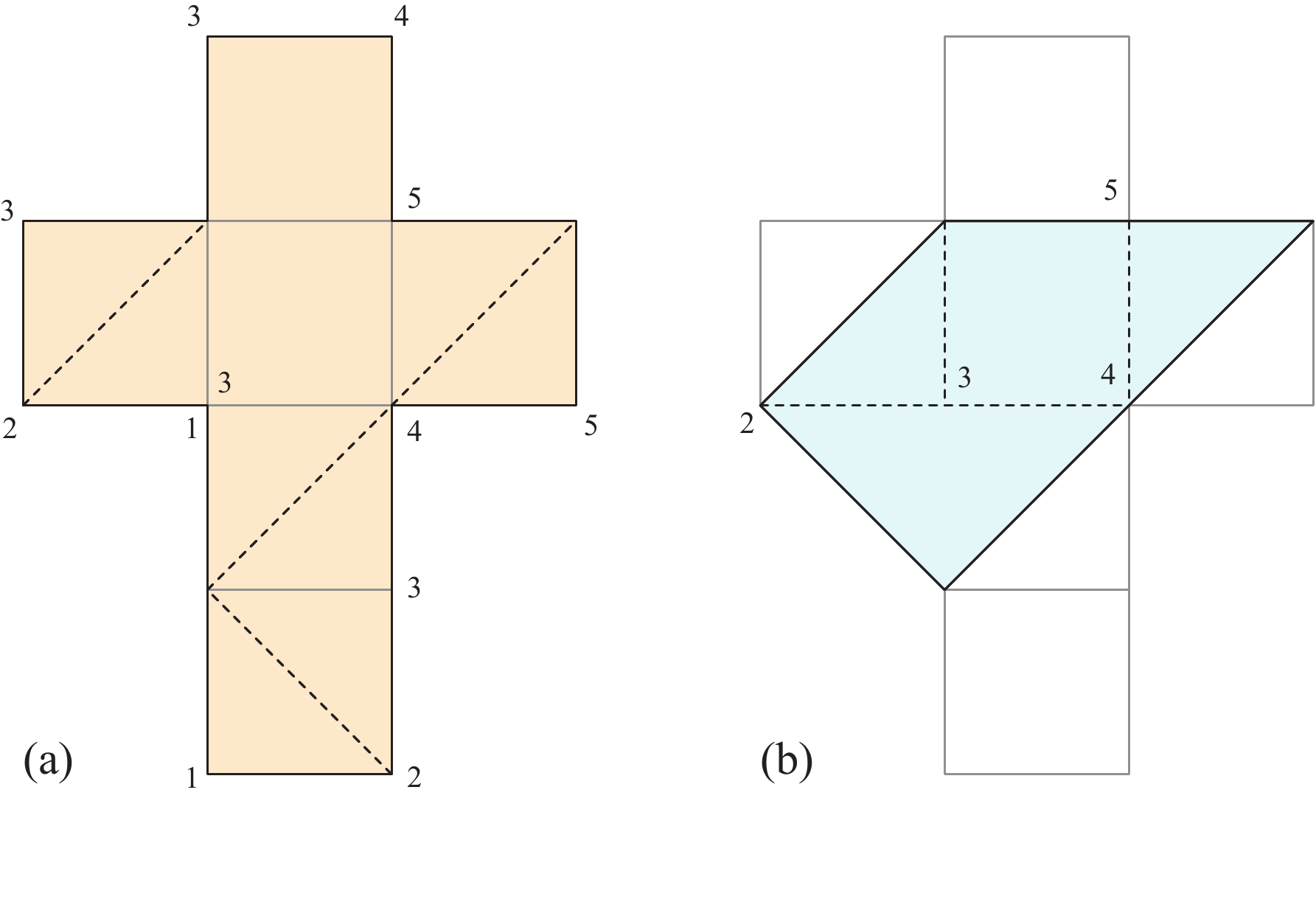}
\caption{Folding the Latin cross cube net to a flat
  quadrilateral polyhedron.
Points with the same label in~(a) are identified in the refolding~(b).}
\figlab{LatinCrossParallelogram}
\end{figure}

Let us say that two polyhedra $Q_1$ and $Q_2$ form a \emph{net pair}
if they may be unfolded to a common polygonal net.
In Figure~\figref{LatinCrossParallelogram}, the cube is cut along
edges to unfold to the Latin cross polygon,
but the flat quadrilateral must have \emph{face cuts} through the interior
of its two faces to unfold to the same Latin cross.

In general there is little understanding of which polyhedra form net
pairs.
See, for example, Open Problem~25.6 in~\cite{do-gfalop-07}.
Here we explore a narrow question on net pairs,
narrow enough to obtain a complete answer.

The cuts to unfold a convex polyhedron to a single polygon
form a spanning tree of the polyhedron's vertices~\cite[Sec.~22.1.3]{do-gfalop-07}.
Shephard explored the special case where the spanning tree is
a Hamiltonian path of the 1-skeleton of the polyhedron,
i.e., all cuts are along polyhedron edges~\cite{s-cpcn-75}.
The result is a \emph{Hamiltonian unfolding} of the polyhedron.
(Note the cube unfolding that produces
Figure~\figref{LatinCrossParallelogram}(a)
is not a Hamiltonian unfolding: the cut tree has four leaves.)
Some combinatorial questions on Hamiltonian unfoldings were explored in~\cite{ddlo-efupp-02};
see~\cite[Fig.~25.59 ]{do-gfalop-07}.
In particular, there are polyhedra that have an exponential number of
combinatorially distinct Hamiltonian
unfoldings: $2^{\Omega(n)}$ for a polyhedron with $n$ vertices.

Another variant is provided by the class of \emph{perimeter-halving} 
foldings~\cite[Sec.~25.1.2 ]{do-gfalop-07},
which correspond to spanning cut paths that may employ face cuts 
rather than solely following polyhedron edges.
In~\cite{lddss-zupc-10} these paths were memorably rechristened as
\emph{zipper paths}, producing \emph{zipper unfoldings}.
We will adopt that nomenclature, including the verbs \emph{zip} and
\emph{unzip}
to mean folding and unfolding (respectively) along zipper paths.
We reserve \emph{Hamiltonian path} to be a zipper path along
polyhedron edges.
Finally, if two polyhedra each unzip to a common polygonal net,
we say they form a \emph{zipper pair}.

The narrow question we explore is this:
\begin{quotation}
\noindent
\textbf{Question:}
Does each of the Platonic solids form a zipper pair
with a flat convex polyhedron,
with the zipper path on the regular polyhedron forming a Hamiltonian
path of its edges?
\end{quotation}

We show that the tetrahedron,\footnote{
    We drop the modifier ``regular'' to shorten the names of the
    five regular polyhedra.
}
the cube, the octahedron, and the
icosahedron all form such zipper pairs with flat parallelogram
polyhedra.
The dodecahedron has no such zipper mate.
Note that it would be too restrictive to insist that both zippers are
Hamiltonian paths of the 1-skeletons, because
for a flat polyhedron, the 1-skeleton is the single cycle bounding
the polygon, and so a Hamiltonian unfolding is just two 
copies of the convex polygon joined along one edge.



\section{Flat Zipper Pairs}
\subsection{Tetrahedron}
The regular tetrahedron has only one Hamiltonian path (up to
symmetries),
which unfolds to the $2 \times 1$ parallelogram
shown in Figure~\figref{TetrahedronParallelogram}(b)
(in Fig.~2 in~\cite{lddss-zupc-10}).
Because this net is a convex polygon,
Thm.~25.1.4 in~\cite{do-gfalop-07} establishes that it has an infinite
number of zippings to various convex polyhedra.
The zipping shown in Figure~\figref{TetrahedronParallelogram}(c)
folds it to a doubly covered rhombus.
\begin{figure}[htbp]
\centering
\includegraphics[width=0.9\linewidth]{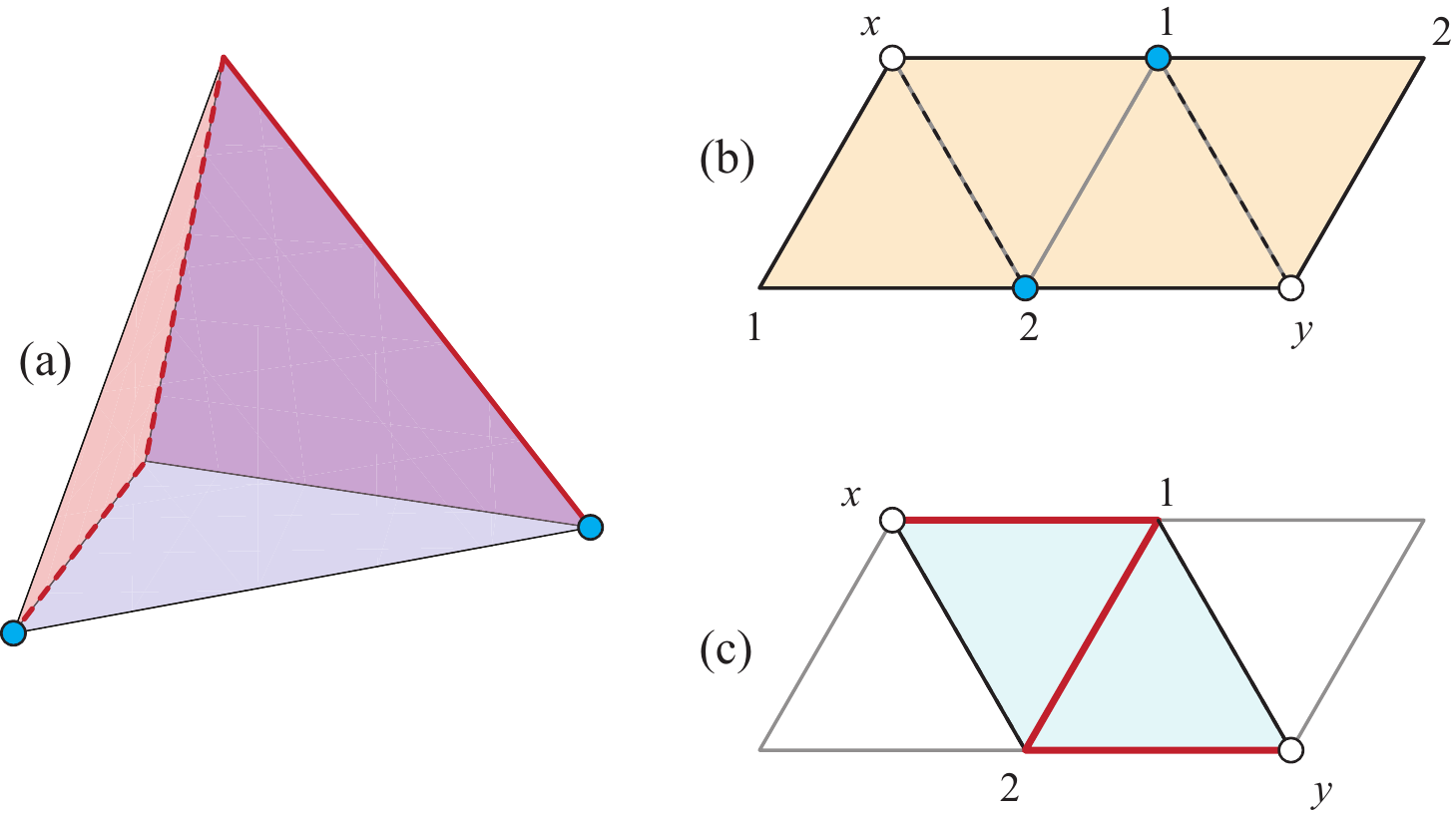}
\caption{(a)~The Hamiltonian cut path on a tetrahedron
leads to the Hamiltonian unfolding~(b),
which zips from $x$ to $y$ (identifying the labeled points) to 
a flat rhombus polyhedron of side length 1~(c).}
\figlab{TetrahedronParallelogram}
\end{figure}

\subsection{Cube}
The cube has three distinct Hamiltonian unfoldings
(Fig.~1 in~\cite{lddss-zupc-10}): one with the path endpoints at
opposite cube corners, and two with the path endpoints at either
end of a cube edge.
One of the latter (shown in Figure~\figref{CubeT}) produces a
`T'-shape
that has no zippings except back to
the cube.
We call such a zipper unfolding \emph{zip-rigid}.
We defer an explanation of how it is known that this
unfolding is zip-rigid to Section~\secref{ZipIt} below.
\begin{figure}[htbp]
\centering
\includegraphics[width=0.8\linewidth]{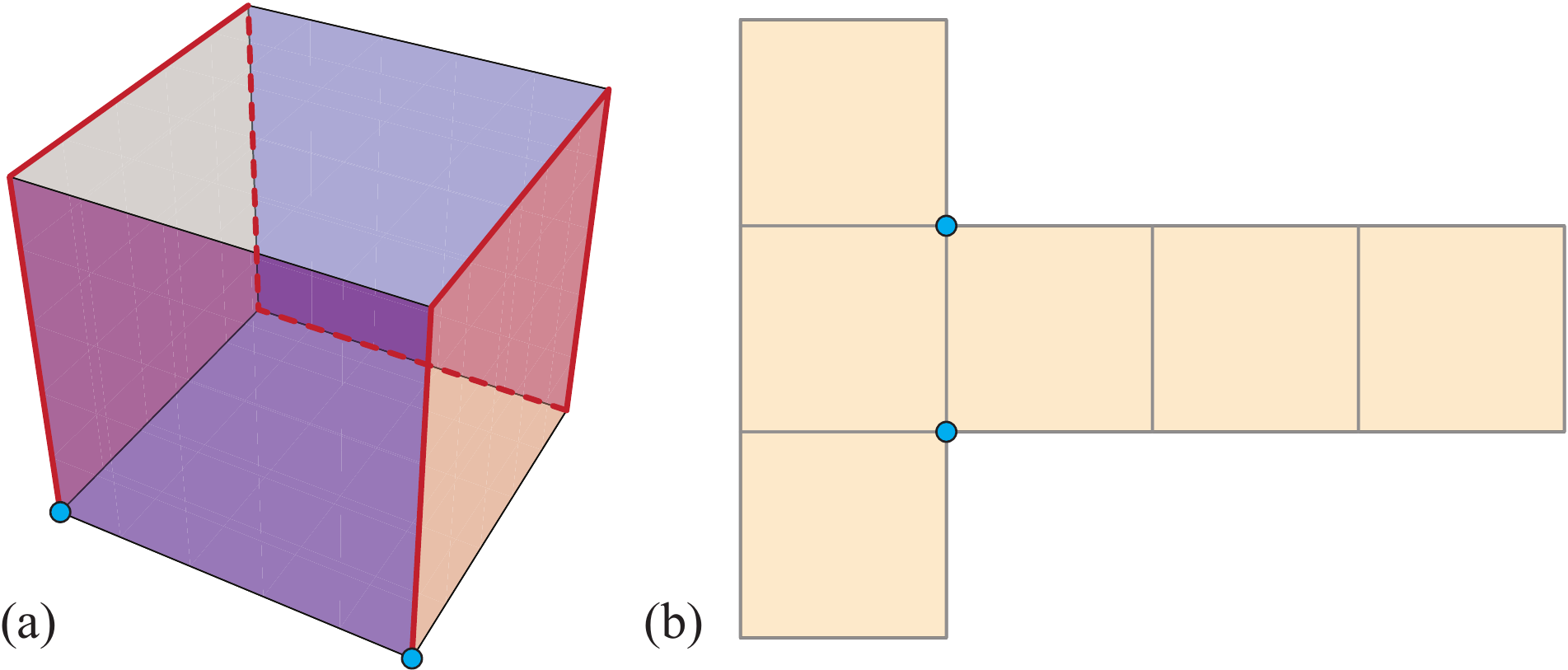}
\caption{(a)~The first Hamiltonian cut path
leads to (b)~a zip-rigid Hamiltonian `T'-unfolding of the cube.}
\figlab{CubeT}
\end{figure}

The other two Hamiltonian unfoldings of the cube,
which we call the `S'- and the `Z'-unfoldings,
both zip to the same
doubly covered parallelogram, as shown in
Figures~\figref{CubeS} and~\figref{CubeZ}.
An animation of the `S'-folding is shown in Figure~\figref{Sanim}.
\begin{figure}[htbp]
\centering
\includegraphics[width=\linewidth]{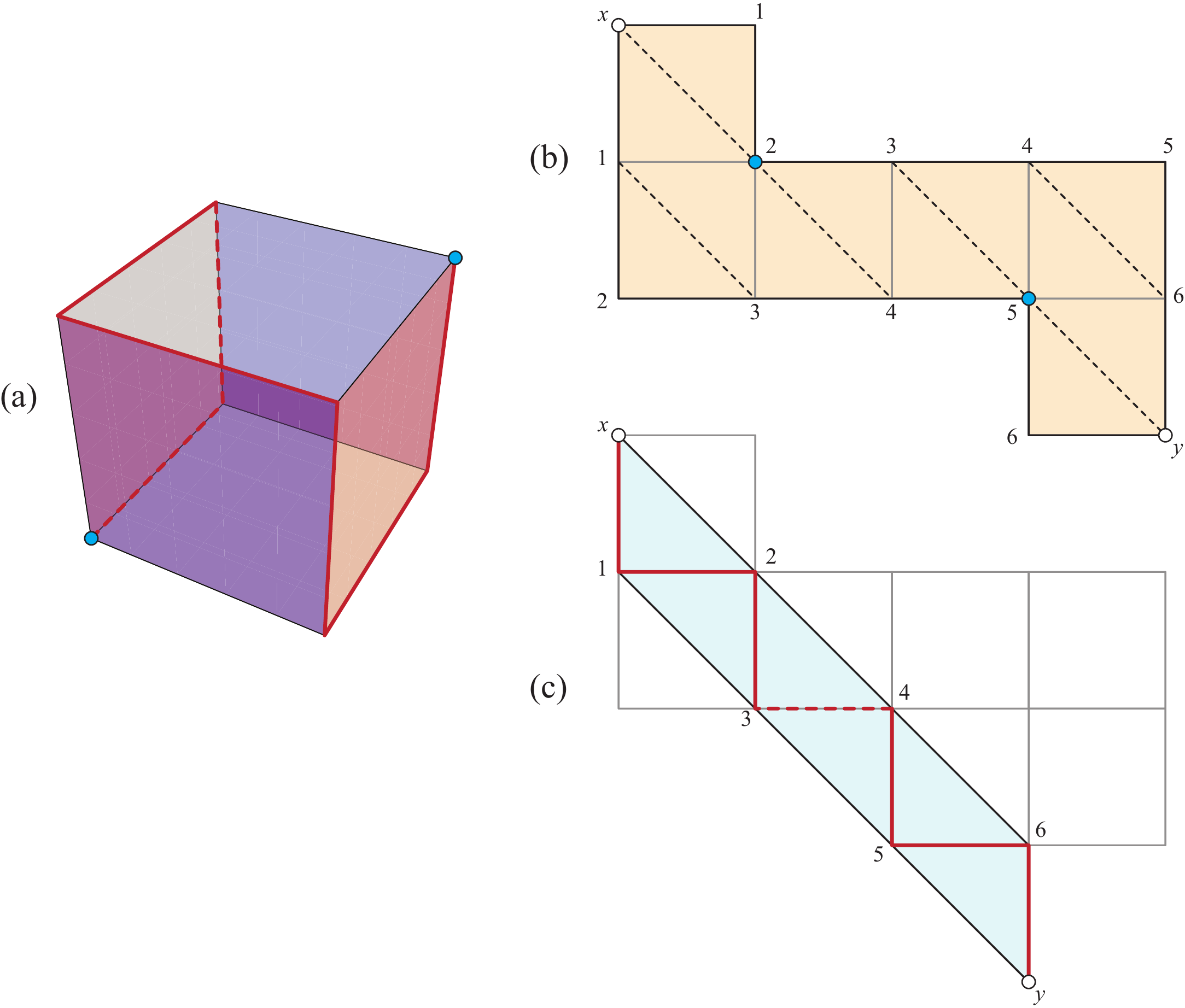}
\caption{(a)~The second Hamiltonian cut path on a cube.
(b)~The resulting Hamiltonian `S' unfolding.
(c)~Zipped according to the indicated point identifications to a
parallelogram polyhedron of side lengths $1$ and $3 \sqrt{2}$.}
\figlab{CubeS}
\end{figure}
\begin{figure}[htbp]
\centering
\includegraphics[width=0.75\linewidth]{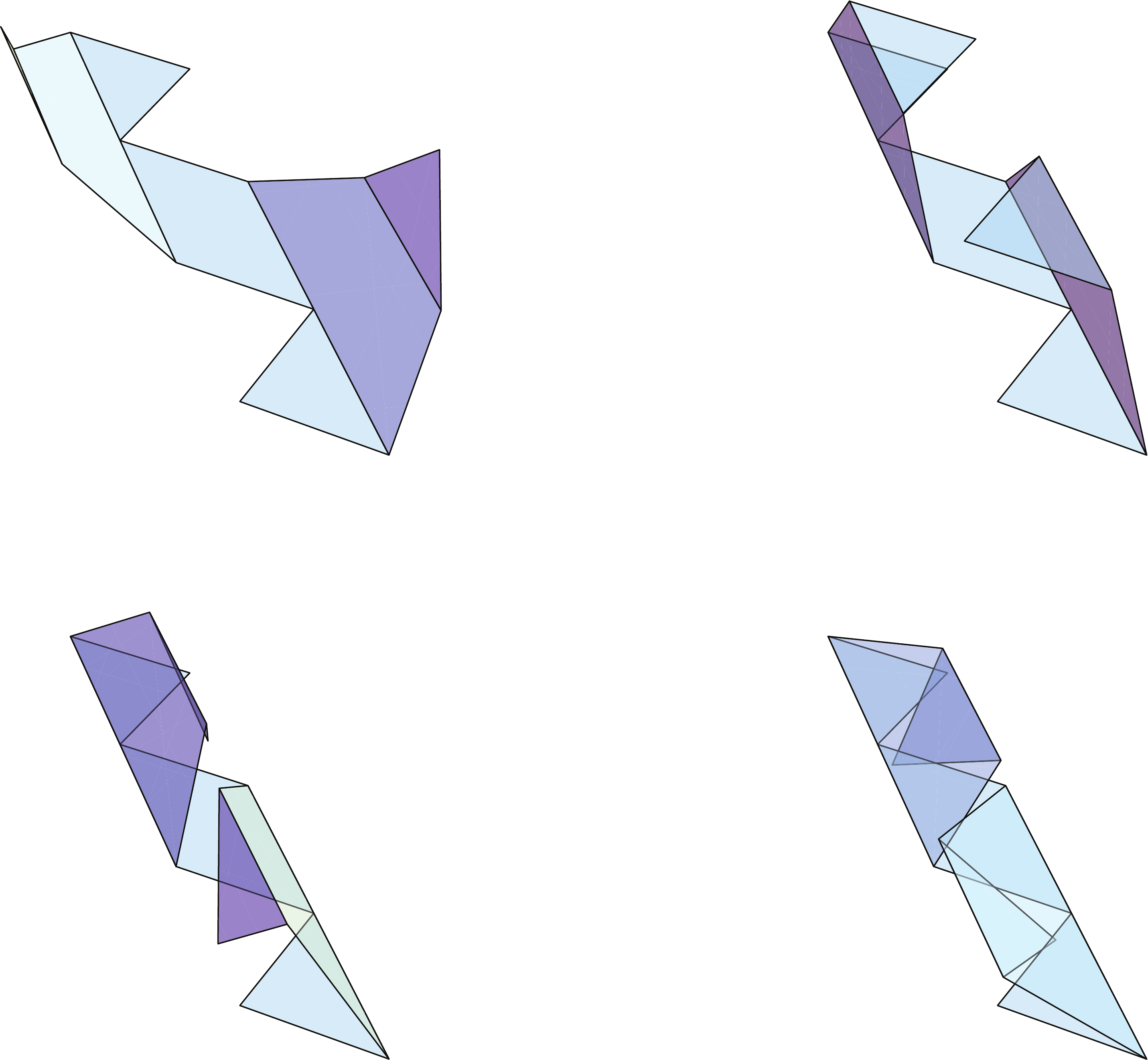}
\caption{Snapshots from an animation folding the parallelogram in
Figure~\protect\figref{CubeS}(b,c).}
\figlab{Sanim}
\end{figure}
\begin{figure}[htbp]
\centering
\includegraphics[width=\linewidth]{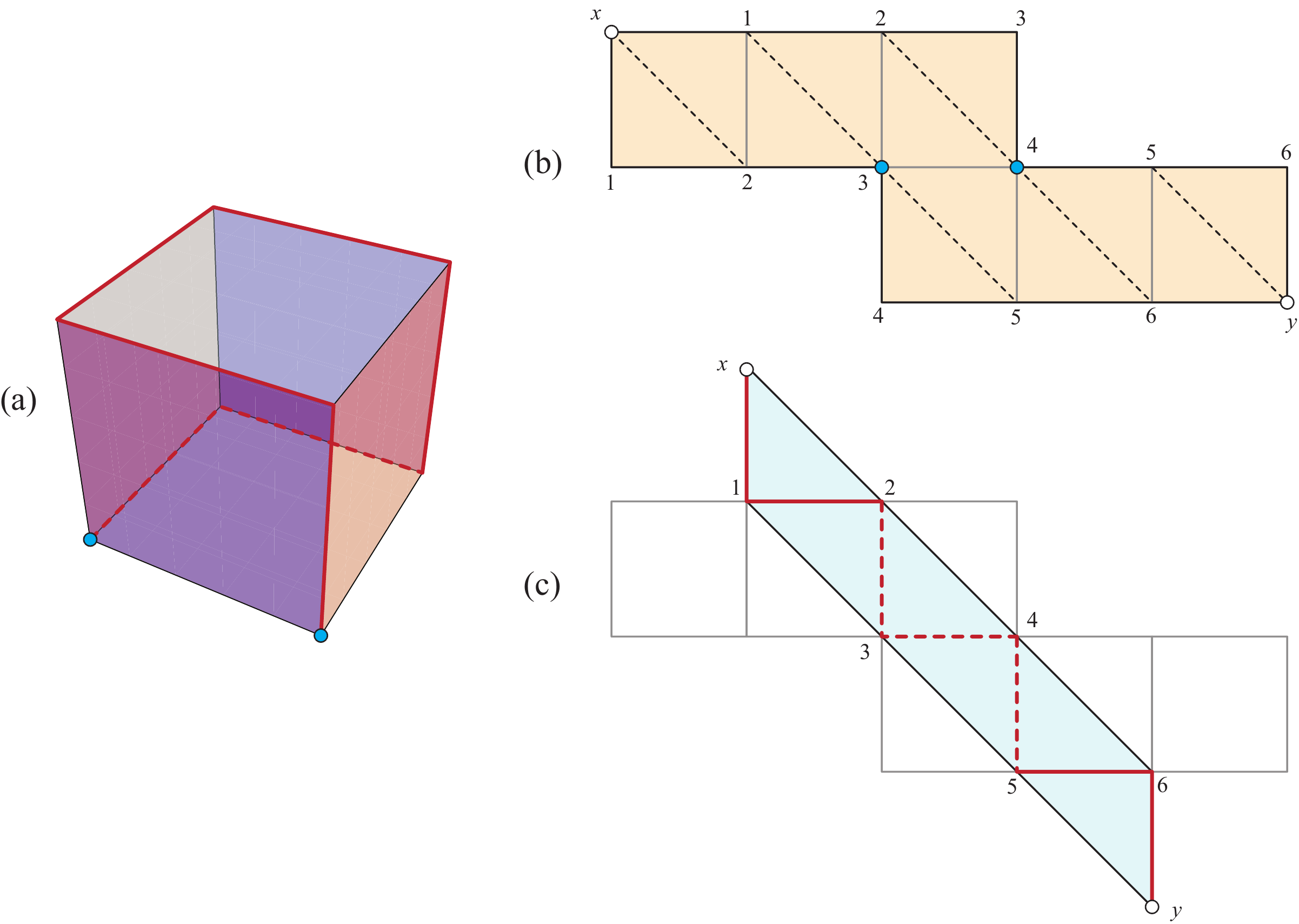}
\caption{(a)~The third  Hamiltonian cut path on a cube.
(b)~The resulting Hamiltonian `Z' unfolding.
(c)~Zipped according to the indicated point identifications to a
parallelogram polyhedron of side lengths $1$ and $3 \sqrt{2}$.}
\figlab{CubeZ}
\end{figure}

\subsection{A Zipping Algorithm}
\seclab{ZipIt}

Let $P$ be a  polygon, the polygonal net corresponding to a zip-pair of
polyhedra $Q_1$ and $Q_2$.
Each of the two zippings of $P$ are perimeter-halving foldings, with
the
endpoints of the zip path bisecting the perimeter.
If we normalize the perimeter of $P$ to 1 and parametrize
it from 0 to 1, we can view the two
zippings
abstractly as in Figure~\figref{ZipCircle}.
One of the zip-path endpoints are at 0 and $\frac{1}{2}$,
and the other zip-path endpoints are at $x$ and $y=x+\frac{1}{2}$.
We seek to find all the locations $x$ that determine a zipping to some
convex polyhedron.

\begin{figure}[htbp]
\centering
\includegraphics[width=0.5\linewidth]{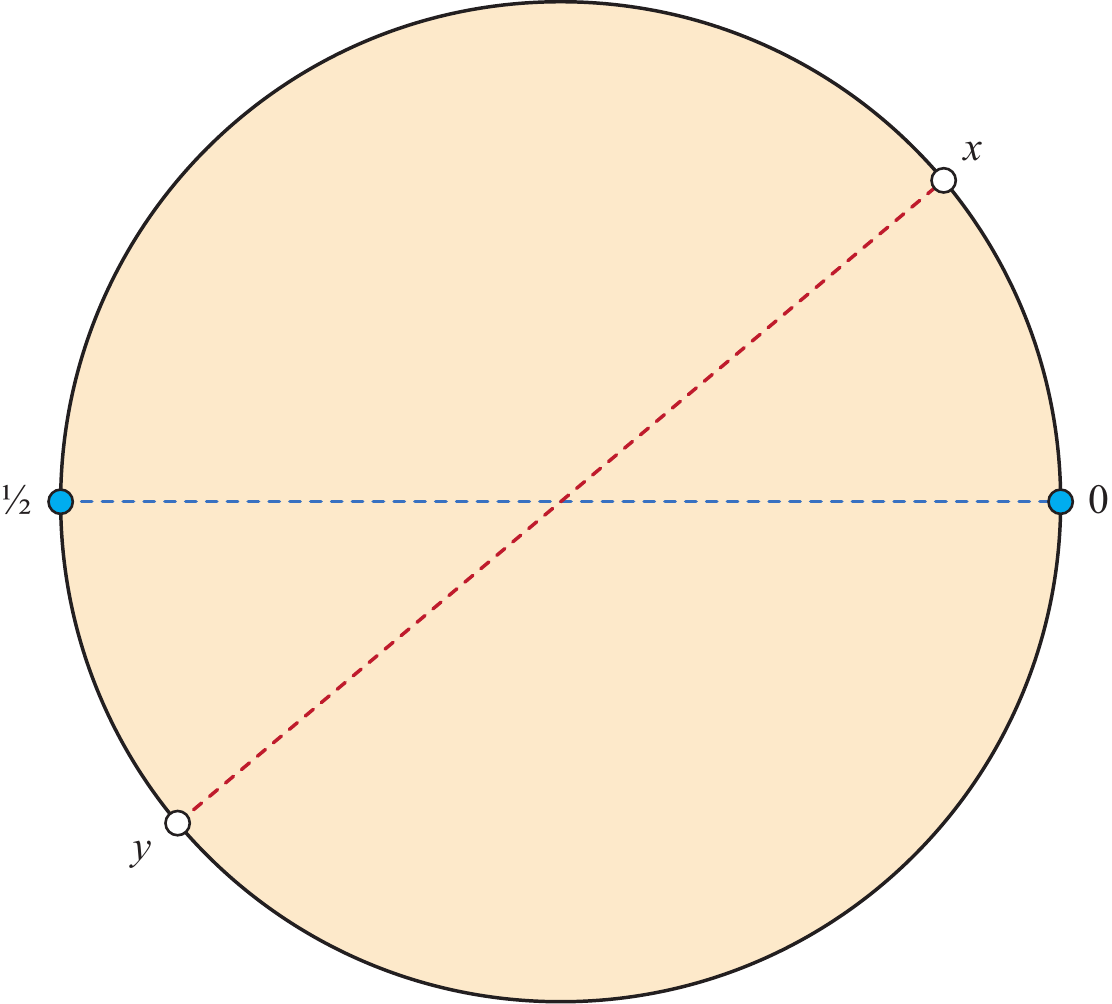}
\caption{A zip-pair, abstractly. The perimeter has been normalized to 1.}
\figlab{ZipCircle}
\end{figure}

As previously mentioned, if $P$ is convex, then every $x$ determines a
convex polyhedron (Thm.~25.1.4 in~\cite{do-gfalop-07}), so we henceforth exclude that case.
If $P$ is not convex, it has at least one reflex vertex $v$ with
internal angle $\b > \pi$.
Now there are only two options at $v$:
(1)~$v$ can serve as $x$, so the zipping starts at $v=x$; or
(2)~Some strictly convex vertex $u_i$ whose internal angle $\a_i$
satisfies $\a_i + \b \le 2 \pi$ is glued to $v$.
If more than one vertex is glued to $v$, then the folding would not be
a zipping, as $v$ would then constitute a junction of degree $> 2$
in the gluing tree (\cite[Sec.~25.3]{do-gfalop-07}).
Note that if $u_i$ glues to $v$, then $x$ is determined: halfway between $u_i$
and $v$ along the perimeter of $P$.
Thus we only need try each $u_i$ in turn, and check that
Alexandrov's conditions hold for the uniquely determined zipping~\cite[Thm.~23.3.1]{do-gfalop-07}).
This incidentally shows that any $P$ with a reflex vertex admits
only $O(n)$ zippings.

For example, applying this algorithm to the cube `Z'-unfolding
in Figure~\figref{CubeZ} results in six zippings: two copies of the
one shown in that figure, two copies of a tetrahedron, one
5-vertex and one 6-vertex polyhedron.

Although this provides a linear-time algorithm for determining all
zippings of $P$, it does not tell us which of these zippings lead
to flat polyhedra.
Although there is an O$(n^3)$ algorithm for deciding if an
Alexandrov gluing is flat~\cite{o-ofpat-10}, this remains
unimplemented.
We resorted to manual folding of the zippings.

\subsection{Octahedron}
The octahedron also has three distinct Hamiltonian paths,\footnote{
   This is natural because the cube and octahedron are duals.
   However, it is shown in~\cite[Fig.~4]{lddss-zupc-10}
   that the dual of a Hamiltonian unfolding is not necessarily
   a Hamiltonian path through the faces of that unfolding.
}
one between
the top and bottom vertices (separated by distance 2 in the
1-skeleton), and two paths between adjacent (distance-1) vertices.
The first Hamiltonian unfolding both zips to a rectangle as shown
in Figure~\figref{OctahedronRect},
and zips to a parallelogram, Figure~\figref{OctahedronSPara}.
I find the rectangle zipping especially surprising, as it derives from
a shape all of whose angles are multiples of $\pi/3=60^\circ$.
\begin{figure}[htbp]
\centering
\includegraphics[width=\linewidth]{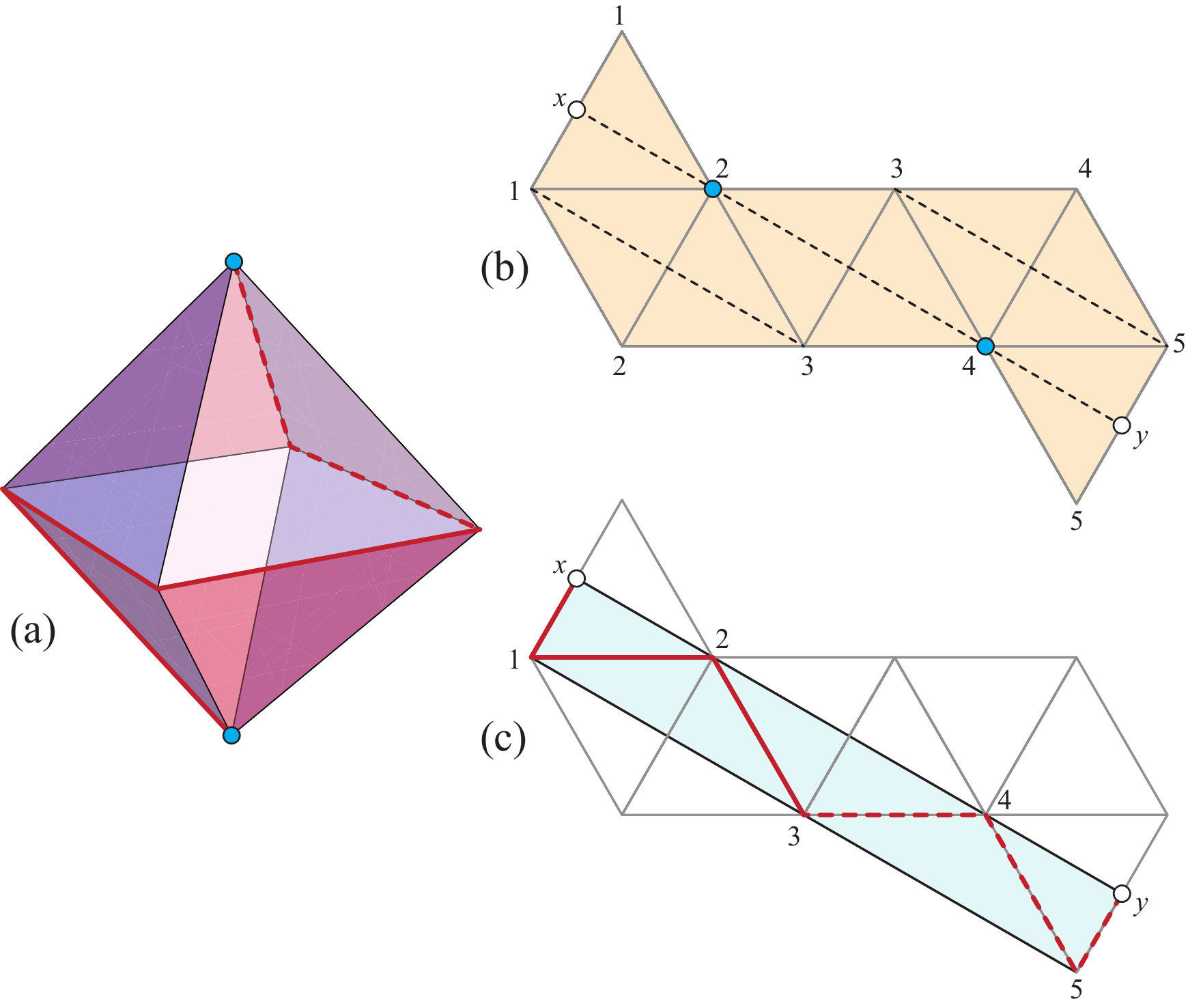}
\caption{(a)~Hamiltonian cut path on an octahedron.
(b)~Its corresponding Hamiltonian unfolding.
(c)~Zipping folds it to
a flat doubly covered rectangle
of dimensions $\frac{1}{2} \times 2\sqrt{3}$.}
\figlab{OctahedronRect}
\end{figure}

\begin{figure}[htbp]
\centering
\includegraphics[width=0.5\linewidth]{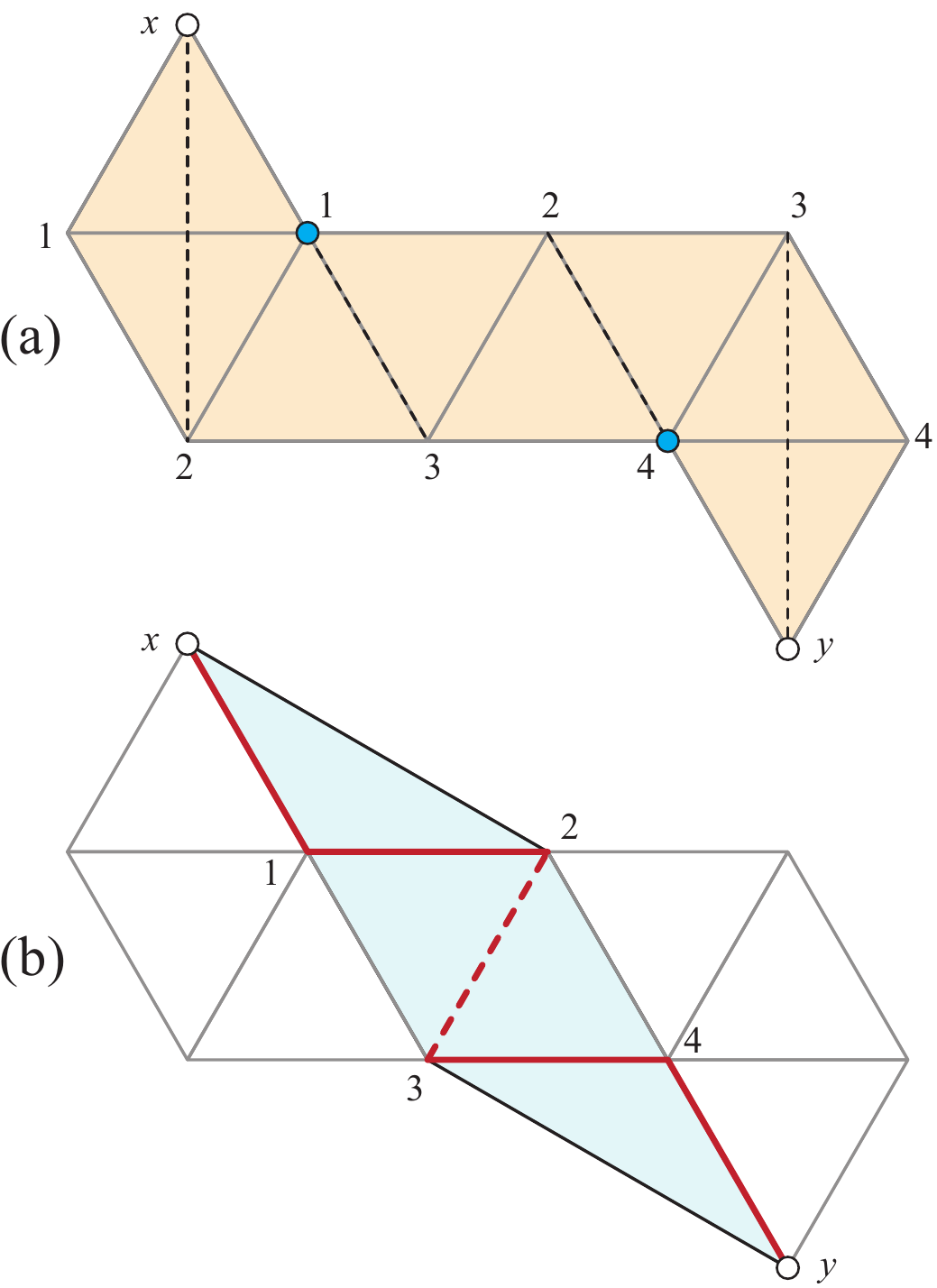}
\caption{(a)~Another zipping of the same unfolding from
  Figure~\protect\figref{OctahedronRect}
leads to (b)~a  $1 \times \sqrt{3}$ parallelogram polyhedron.}
\figlab{OctahedronSPara}
\end{figure}

One of the other Hamiltonian unfoldings of the octahedron,
shown in Figure~\figref{OctahedronZPara},
zips to a parallelogram.
The other Hamiltonian unfolding does not have a flat zipping,
although it does have zippings, e.g., to a tetrahedron all four of
whose
vertices have curvature $\pi$.

\begin{figure}[htbp]
\centering
\includegraphics[width=\linewidth]{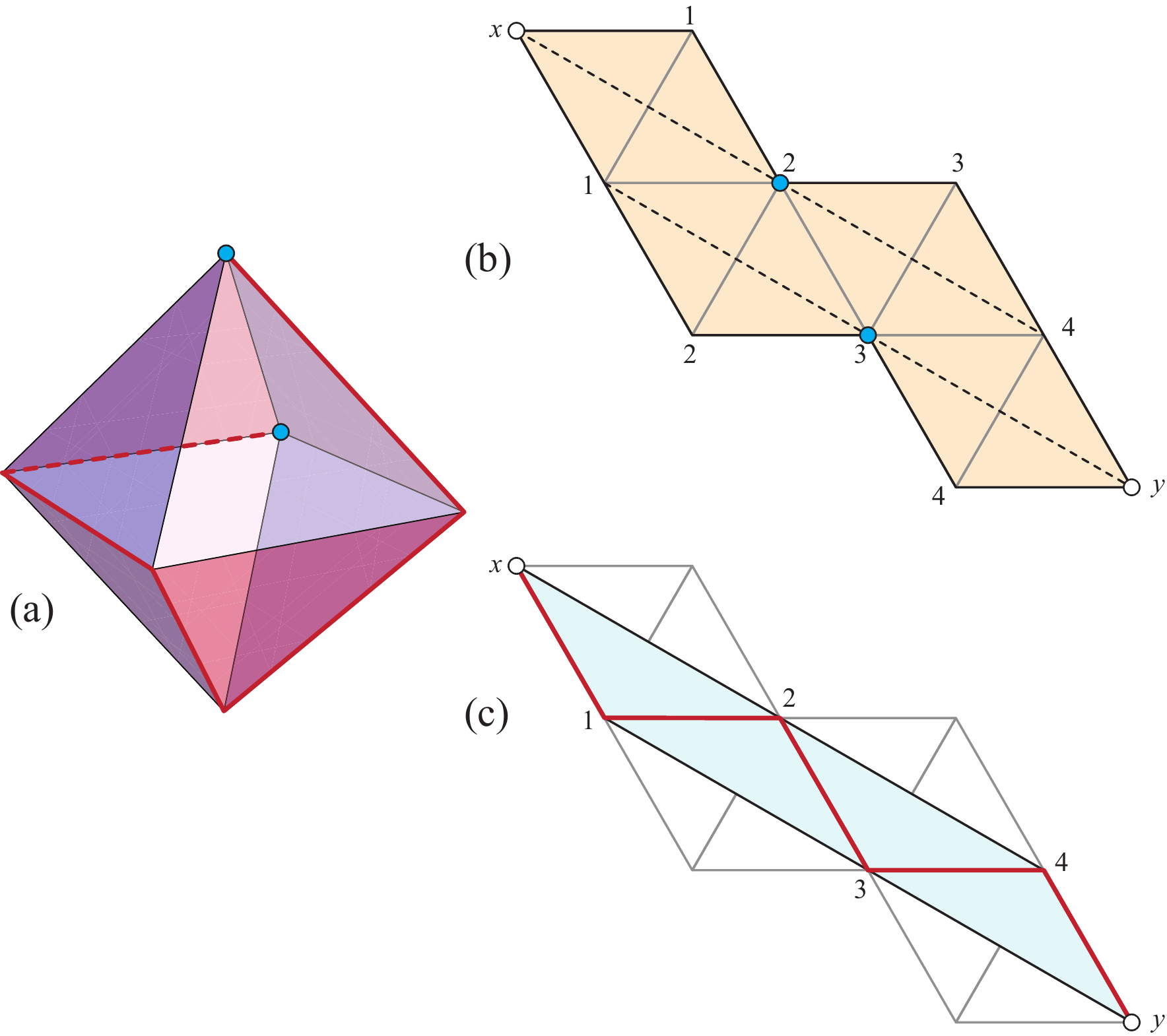}
\caption{(a)~Hamiltonian cut path on an octahedron.
(b)~Its corresponding Hamiltonian unfolding.
(c)~Zipping folds it to
a flat doubly covered parallelogram
of dimensions $1 \times 2\sqrt{3}$.}
\figlab{OctahedronZPara}
\end{figure}

I cannot resist mentioning that this last net folds to a flat rectangular
polyhedron, whose cut tree, however, is not a zipping:
Figure~\figref{OctahedronNonZip}.
\begin{figure}[htbp]
\centering
\includegraphics[width=0.5\linewidth]{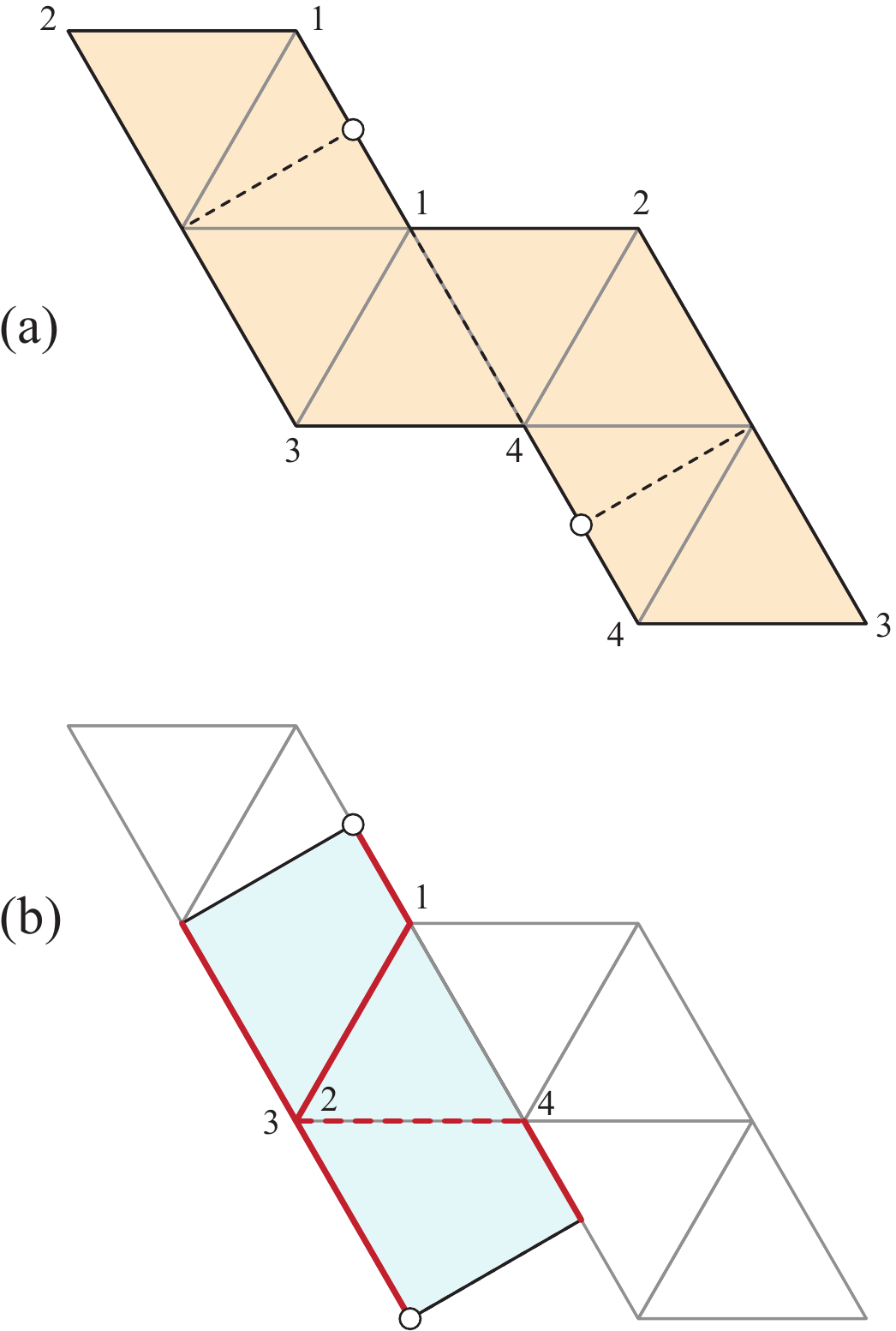}
\caption{The same Hamiltonian unfolding from
  Figure~\protect\figref{OctahedronZPara}
folds to a $\frac{\sqrt{3}}{2} \times 2$ doubly covered rectangle,
but this folding is not a zipping.}
\figlab{OctahedronNonZip}
\end{figure}

\subsection{Dodecahedron}
Every Hamiltonian unfolding of the dodecahedron is zip-rigid,
and therefore it has no flat zip pair in the sense posed in our
Question above.
The reason is as follows.
Let $x$ and $y$ be the endpoints of the Hamiltonian path that unfolds
the dodecahedron.  Then the reflex angle of the net at $x$ and $y$
is $3 \cdot \frac{3}{5} \pi = 324^\circ$, leaving an external angle of
$36^\circ$ there.
The smallest convex angle in any edge unfolding of the dodecahedron
is $\frac{3}{5} \pi =108^\circ$, so no vertex can glue into $x$ or $y$.
Therefore, a zipping must zip at $x$ and $y$, leading directly back
to the dodecahedron.

We should mention that loosening the criteria posed in our Question
leads to a flat refolding of a Hamiltonian net for the dodecahedron.
Figure~\figref{DodecahedronFlat} illustrates one such,
using the unfolding
in Fig.~2 in~\cite{lddss-zupc-10}.
Here the refolding in Figure~\figref{DodecahedronFlat}(c) is 
neither convex nor a zipper folding.
\begin{figure}[htbp]
\centering
\includegraphics[width=\linewidth]{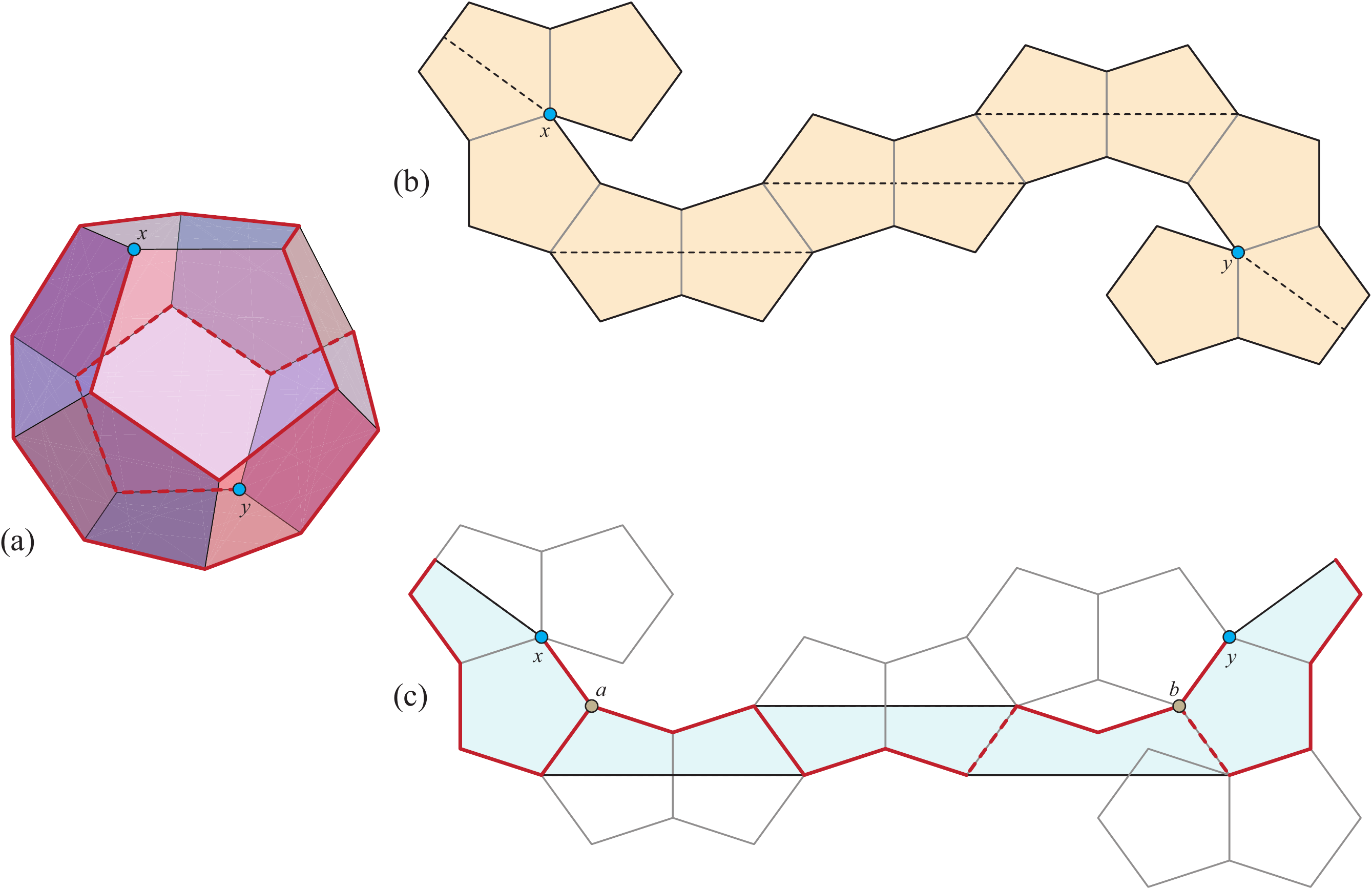}
\caption{(a)~Hamiltonian cut path on a dodecahedron.
(b)~Its corresponding Hamiltonian unfolding.
(c)~A non-zipper refolding to a doubly covered flat nonconvex polygon.
The cut tree has degree 3 at vertices $a$ and $b$.}
\figlab{DodecahedronFlat}
\end{figure}

\subsection{Icosahedron}
For the tetrahedron, cube, and octahedron, it was easy to explore
all the Hamiltonian unfoldings, because there are so few (1, 3, and 3
respectively).
The icosahedron, however, has hundreds of Hamiltonian unfoldings.
At this writing, I do not know precisely how many geometrically
distinct Hamiltonian unfoldings it possesses.

The diameter of the icosahedral graph is 3, so the end points of a
Hamiltonian path are a distance 1, 2, or 3 apart.
Fixing two vertices separated by a distance $d \in \{1,2,3\}$,
I found 
that there are, respectively,
512, 608, and 720 \emph{labeled} Hamiltonian paths between them.\footnote{
  It is a curious fact that the number of labeled Hamiltonian cycles
  through any fixed edge is $2^9 = 512$.
  The simplicity of this expression suggests there might be 
  a combinatorial explanation,
  a question I asked on MathOverflow,
  \url{http://mathoverflow.net/questions/37788/}.
}
Of course not all these labeled paths are distinct geometric paths
because of symmetries.
However, I have not carried out the more difficult enumeration of the 
number of geometrically distinct (incongruent as paths in $\R^3$)
Hamiltonian paths on an icosahedron.
But certainly this number is less than $512+608+720 = 1840$.

For each of these 1840 Hamiltonian unfoldings, I ran the zipping
algorithm
in Section~\secref{ZipIt}, which determined that 82 of the unfoldings
had at least one zipping, while all the others are zip-rigid
($82 = 12 + 20 + 50$ in the three classes, respectively).
By visual inspection, 18 of these Hamiltonian
unfoldings
are distinct; they are displayed in 
Figure~\figref{IcosaUnfoldingsDistinct}.
\begin{figure}[htbp]
\centering
\includegraphics[width=\linewidth]{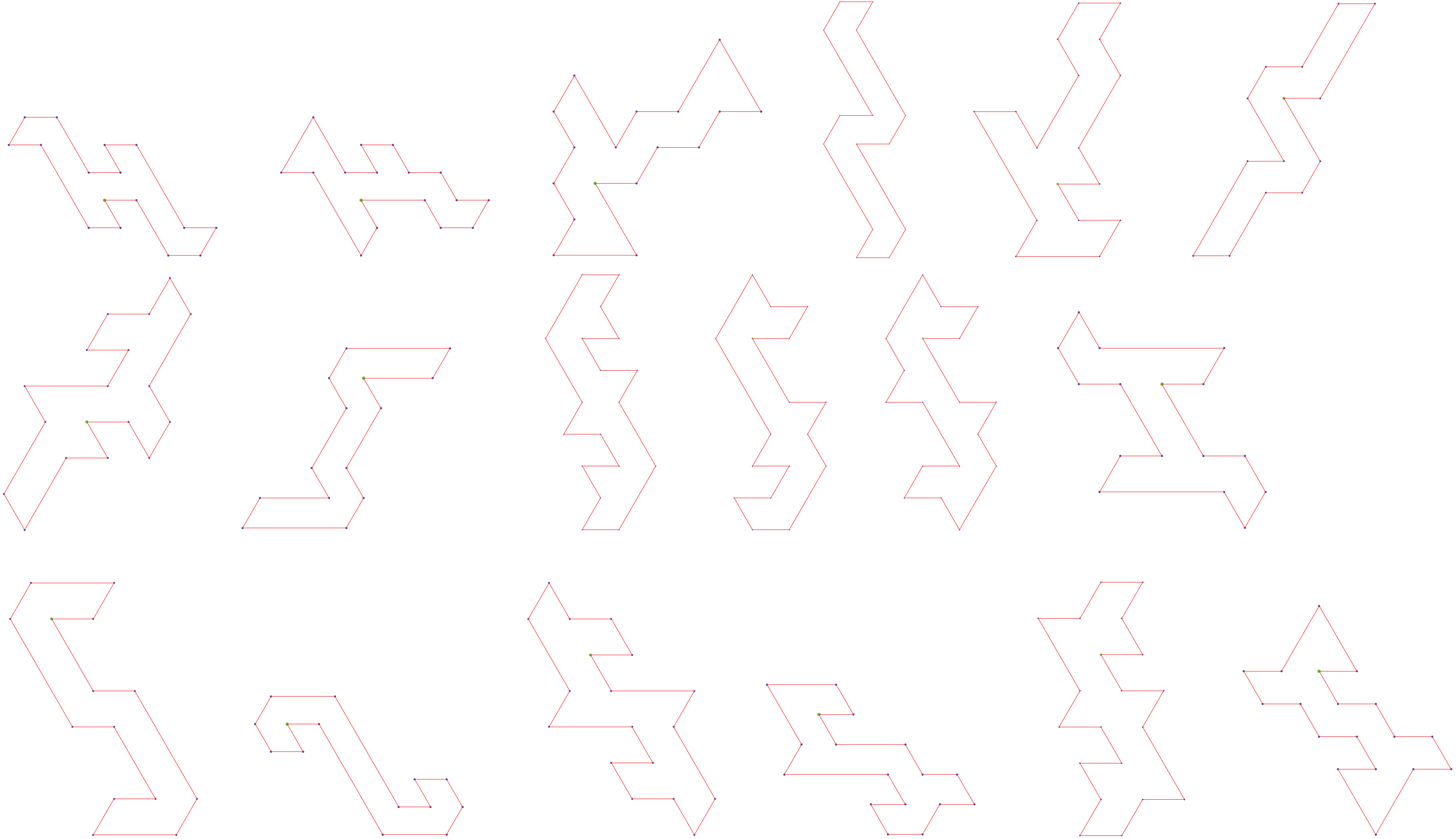}
\caption{The 18 distinct Hamiltonian unfolding of the icosahedron that
each have at least one zipping to another convex polyhedron
(Not all are displayed to the same scale.)}
\figlab{IcosaUnfoldingsDistinct}
\end{figure}

Exactly one of these unfoldings (the leftmost
in the first row) zips to a parallelogram, as 
shown in 
Figure~\figref{IcosahedronParallelogram}.
\begin{figure}[htbp]
\centering
\includegraphics[width=\linewidth]{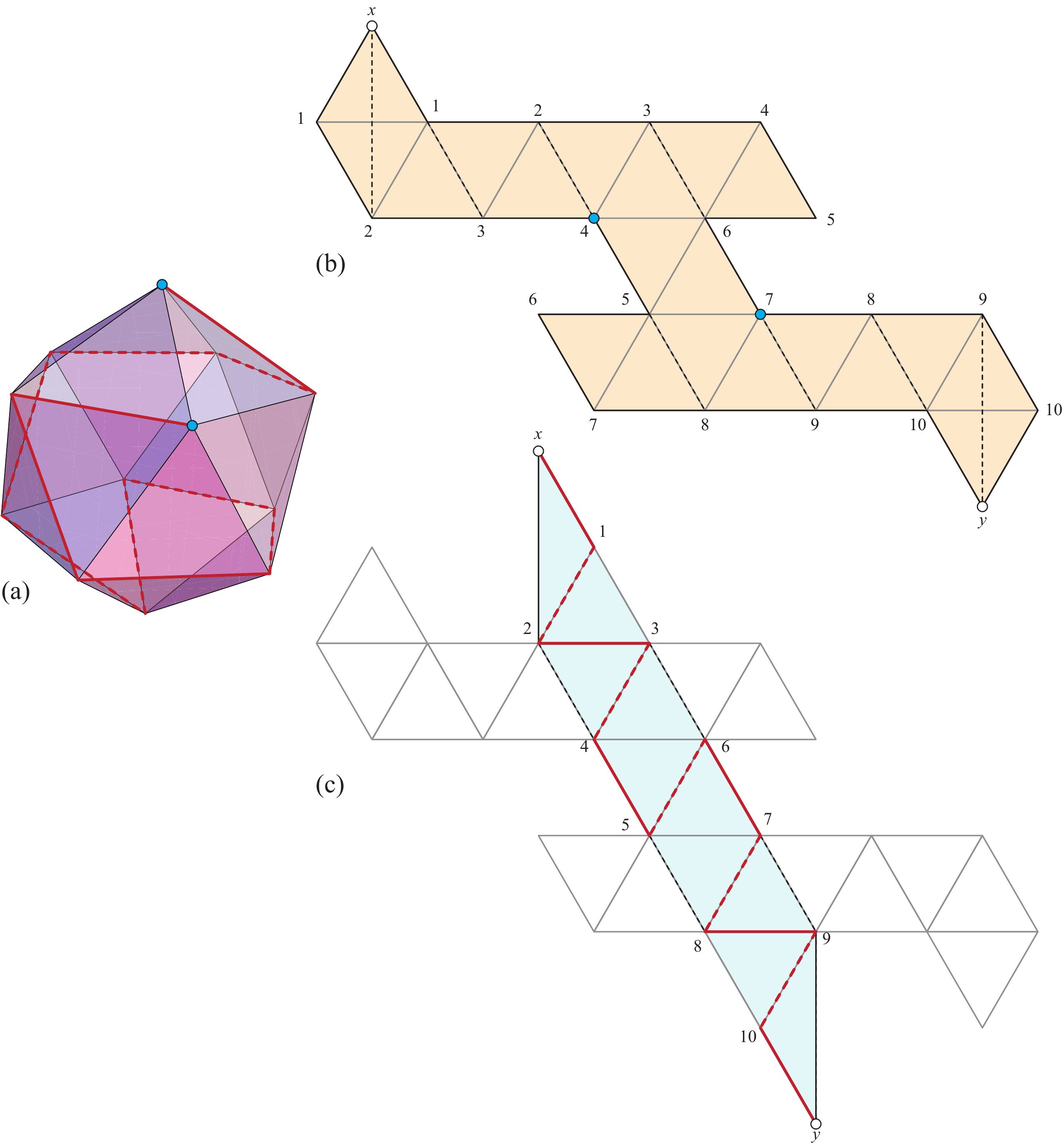}
\caption{(a)~Hamiltonian cut path on an icosahedron.
(b)~Its corresponding Hamiltonian unfolding.
(c)~Rezipping folds it to
a flat doubly covered parallelogram
of side lengths $\sqrt{3}$ and 5.}
\figlab{IcosahedronParallelogram}
\end{figure}
None of the other zips to a parallelogram 
(but it remains
possible that there is some zipping to flat polyhedron different
from a doubly covered parallelogram).

\section{Future Work}
\seclab{Future}
As is evident from the foregoing, there is little theory behind the
unfoldings detailed here.
The central open problem is to gain more insight into which polyhedra
are net pairs, or more specifically, zipper pairs.
Perhaps intuition can be strengthened by tackling specific subquestions that
fall under this general umbrella.  It is easy to list such questions,
all of are open because of the lack of a general theory.
For example, the Hamiltonian unfoldings of the Archimedean solids
detailed in~\cite{lddss-zupc-10} could be explored.

An interesting specific but tangential question raised by this work is to determine the exact
number of geometrically distinct Hamiltonian paths on a regular icosahedron.

\paragraph{Acknowledgments.}
I thank Stephanie Annessi and Katherine Lipow for help in enumerating
and folding the
icosahedron Hamiltonian unfoldings.

\bibliographystyle{alpha}
\bibliography{/Users/orourke/bib/geom/geom}
\end{document}